\documentclass[aps,prl,twocolumn,showpacs]{revtex4}
\usepackage{graphicx}
\usepackage{color}
\usepackage{amssymb}
\def\cN{{\cal N}}
\def\vv{{\bf v}}
\def\vp{{\bf p}}
\def\vz{{\bf z}}
\def\vR{{\bf R}}
\def\Tr#1{\mbox{Tr}\left(#1\right)}
\def\pder#1#2{\mbox{$\displaystyle\frac{\partial #1}{\partial #2}$}}
\newcommand{\D}{\mathfrak{D}}

\newcommand{\g}{\mathfrak{g}}
\newcommand{\hG}{\hat\mathfrak{G}}

\newcommand{\Area}{{\cal A}}
\newcommand{\tinyonehalf}{\frac{\mbox{\tiny 1}}{\mbox{\tiny 2}}}

\newcommand{\sech}{\mbox{sech}}
\begin{document}
\title{Phase Modulated Thermal Conductance of Josephson Weak Links}
\author{Erhai Zhao, Tomas L$\mathrm{\ddot{o}}$fwander and J. A. Sauls}
\affiliation{Department of Physics \& Astronomy, Northwestern University, Evanston, IL 60208}
\date{\today}
\begin{abstract}
We present a theory for quasiparticle heat transport through superconducting weak links. The
thermal conductance depends on the phase difference ($\phi$) of the superconducting leads. Branch conversion
processes, low-energy Andreev bound states near the contact and the suppression of the local
density of states near the gap edge are related to phase-sensitive transport processes. Theoretical
results for the influence of junction transparency, temperature and disorder,
on the phase modulation of the conductance are reported.
For high-transmission weak links, $D\rightarrow 1$, the formation of an Andreev bound state at
$\epsilon_{\text{\tiny b}}=\Delta\cos(\phi/2)$ leads to suppression of the density of states
for the continuum excitations that transport heat, and thus, to a reduction in the conductance for
$\phi\simeq\pi$.
For low-transmission ($D\ll 1$) barriers resonant scattering at energies
$\epsilon\simeq(1+D/2)\Delta$ leads to an increase
in the thermal conductance as $T$ drops below $T_c$ (for phase differences near $\phi=\pi$).
\end{abstract}
 \maketitle
The Josephson effect in superconducting weak links is perhaps the best known example of macroscopic phase
coherence. In addition to the superconducting tunnelling current, $j_s=j_c \sin \phi$, Josephson
\cite{jos62} showed that the electrical current through a tunnel junction includes Ohmic terms,
$j_d=(\sigma_0+\sigma_1\cos\phi)V$, where $\phi$ is the phase difference between the two
superconductors, and $V$ is the voltage across the junction. These terms describe dissipative quasiparticle
tunnelling when the junction is biased by a voltage. The term, $\sigma_{1}V\cos\phi$, is attributed
to ``interference'' between Cooper-pair and quasiparticle tunnelling \cite{jos62,har74,lan74},
and averages to zero in voltage-biased junctions.

Phase-modulated dissipative currents are characteristic of any type
of superconducting weak link. For example, the thermal current through a temperature-biased SIS junction is
predicted to be a periodic function of $\phi$ \cite{mak65}. For a stationary phase difference the
thermal conductance is also stationary. Thus, in contrast to voltage-biased junctions, the phase-modulated
thermal conductance does not average to zero. However, less is known about phase-sensitive thermal transport
across superconducting weak links compared with their current-voltage characteristics. Recent investigations
of heat transport through SIS junctions are based on the tunnel Hamiltonian method \cite{gut97}, while
Kulik and Omel'yanchuk \cite{kul92} calculated the thermal current for the opposite extreme of a perfectly
transmitting superconducting constriction (a ``pinhole''). In terms of the transmission coefficient of the
interface potential barrier between two superconductors, the SIS junction corresponds to
$D\ll 1$, while the pinhole corresponds to perfect transmission, $D\rightarrow 1$.

In the following we present a theory for quasiparticle heat transport through superconducting point contacts
for any junction transparency, and as a function of temperature and disorder. The thermal
conductance is sensitive to spatial inhomogeneities of the order parameter, particularly changes in phase,
which lead to branch conversion between particle- and hole-like excitations \cite{and64}, and to the formation
of low-energy bound states in the vicinity of the point contact. The bound state spectrum and transmission
probabilities for continuum excitations are strongly modified by the junction transparency, which
leads to large changes in the thermal conductance of the junction.

To study quasiparticle transport through temperature-biased superconducting weak links, we use the method of
nonequilibrium quasiclassical Green functions \cite{esc00}. In this formalism the advanced and retarded Green
functions, $\hG^{A,R}$, describe the local spectrum of excitations for the system, while the Keldysh Green
functions, $\hG^K$, carry the information about the nonequilibrium population of these states. Each
propagator, $\hG^{A,R,K}(\vp_f,\epsilon;\vR,t)$ is a $2\times 2$ matrix in particle-hole (Nambu) space obeying
transport-like equations for excitations of energy $\epsilon$ moving along classical trajectories labelled by
the Fermi momentum, $\vp_f$. We use the notation of Ref. \cite{esc00}, set $\hbar=1, k_B=1$, and consider
spin-independent transport in spin-singlet superconductors.

\begin{figure}
\includegraphics[height=2in]{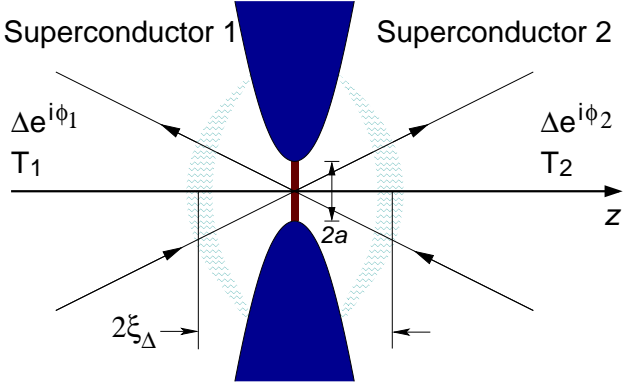}
\caption{Temperature-biased ScS weak link geometry. Quasiparticle trajectories are
         coupled by reflection and transmission at $z=0$.
         The shaded boundaries define the region ($\sim\xi_{\mbox{\tiny $\Delta$}}\gg a$)
         where superconductivity and the excitation spectrum are strongly modified for $\phi_1\ne\phi_2$.}
\label{fig:geometry}
\end{figure}

The basic model for superconducting weak links considered here is that of a constriction of diameter and length on the
nanometer scale, much smaller than the coherence length,
$\xi_{\mbox{\tiny $\Delta$}}=v_f/\pi\Delta$, and the bulk elastic and
inelastic mean free paths (see Fig. \ref{fig:geometry}). The potential barrier located at $z=0$ is characterized by
a transmission (or reflection) probability, $D(\vp_f)$ (or $R(\vp_f)=1-D(\vp_f)$), for normal-state quasiparticles with
Fermi momentum $\vp_f$ incident on the interface. The coupling between the two superconductors, $S_1$ and $S_2$, can
then be described by a boundary condition connecting the Green functions for the two superconductors at the junction
interface \cite{zai84}. The order parameter at the junction interface for the superconductor $S_j$ is
$\Delta\,e^{i\phi_j}$, and the temperature at the junction interface for superconductor $S_j$ is $T_j$ for
$j=1,2$. The phase difference is denoted by $\phi=\phi_2-\phi_1$, and the temperature bias is
$\delta T=T_2-T_1$.

Recently, Eschrig \cite{esc00} recast Zaitsev's boundary condition into a convenient form by parameterizing
the quasiclassical Green functions and transport equations using Shelankov's projection operators \cite{she80}
and generalized spectral, $\gamma^{R,A}$, and distribution, $x^K$, functions obeying Riccatti-type transport
equations. We use the Riccatti formulation of the quasiclassical equations with the boundary condition of Ref.
\cite{esc00} to solve for the Keldysh Green's functions and heat current for superconducting weak links driven
out of equilibrium by a temperature bias. The propagators for trajectories incident,
$\hG_{+}^{R,A,K}=\hG_{1}^{R,A,K}(\vp_f\cdot \hat{\vz}>0, z=0^-)$, and reflected,
$\hG_{-}^{R,A,K}=\hG_{1}^{R,A,K}(\vp_f\cdot \hat{\vz}<0, z=0^-)$ by the interface from the $S_1$ side can be
used to determine the heat current through the interface,
\begin{equation}
\hspace*{-2mm}I_{\epsilon}=\Area N_f\int\frac{d\epsilon}{4\pi i}\epsilon v_f
             \Big\langle\Tr{\hG^K_+(\vp_f,\epsilon)-\hG^K_-(\vp_f,\epsilon)}\Big\rangle\,,
\label{eq:current}
\end{equation}
where $N_f$ is the normal-state density of states at Fermi surface, $\vv_f$ is the
Fermi velocity and $\Area=\pi a^2$ is the
cross-sectional area of the constriction. The Fermi-surface average includes the direction
cosine for projection of the group velocity along the direction normal to the interface; i.e.
$\langle\ldots\rangle={1\over 2}\int_0^{\pi/2}d\theta\sin\theta\cos\theta(\ldots)$, where
$\theta=\arccos(\hat{\vv}_f\cdot\hat{\vz})$.
We consider the case in which $S_1$ and $S_2$ are identical $s$-wave superconductors with isotropic Fermi
surfaces; it is straightforward to generalize the results to two different s-wave superconductors. We also
consider small junctions, $a\ll\xi_{\mbox{\tiny $\Delta$}}$; in this limit the Josephson supercurrent through
the contact is small by at least a factor $a/\xi_{\mbox{\tiny $\Delta$}}$ compared to the bulk critical current, so
pair-breaking corrections to the order parameter by the supercurrent can be neglected \cite{kul78}. Thus, to a
good approximation the order parameters take their bulk equilibrium values, but with local values for the
phase. Also for small constrictions the thermal resistance of the junction is much larger than in that in the
bulk, so the temperature drop occurs essentially at the junction.

With these considerations, $\hG^K_{\pm}$ can be calculated as follows.
First, we express the Green function in terms of Riccati amplitudes \cite{esc00}. For example,
\begin{equation}
\hG^K_{+}=\frac{-2\pi i}{N_1^RN_1^A}
\left(
\begin{array}{cc}
x_1^K+\tilde X_1^K\gamma_1^R\tilde\gamma_1^A & x_1^K\Gamma_1^A-\tilde X_1^K\gamma_1^R\\
x_1^K\tilde\Gamma_1^R-\tilde X_1^K\tilde\gamma_1^A & \tilde X_1^K+x_1^K\tilde\Gamma_1^R\Gamma_1^A
\end{array}
\right),\label{GreenK}
\end{equation}
where
$N_1^RN_1^A=(1+\gamma_1^R\tilde\Gamma_1^R)(1+\tilde\gamma_1^A\Gamma_1^A)$. The advanced
amplitudes are obtained from retarded functions using the
symmetry $\gamma^A=-(\tilde\gamma^R)^*$. In the point contact
geometry, the amplitudes and distribution functions for incoming quantities (lower case)
take their local equilibrium
values: $\gamma^{R}_j=\tilde{\gamma}^R_j(-\epsilon)^*=
-i\Delta\,e^{i\phi_j}/(\varepsilon^R+i\sqrt{\Delta^2-(\varepsilon^R)^2})$,
$x^K_j=\tilde{x}^K_j(-\epsilon)^*=(1-|\gamma_j^{R}|^2)\tanh(\epsilon/2T_j)$,
for $j=1,2$. Using the boundary conditions of Ref. \cite{esc00}, we
construct the corresponding functions for outgoing trajectories (upper case),
\begin{eqnarray}
\tilde\Gamma_1 &=& \frac{\mbox{\small$R(1+\tilde\gamma_2\gamma_2)\tilde\gamma_1+D(1+\tilde\gamma_1\gamma_2)\tilde\gamma_2$}}
                        {\displaystyle{1+R\gamma_2\tilde\gamma_2+D\tilde\gamma_1\gamma_2}}
\,,
\\
\tilde{X}_1    &=& \frac{\mbox{\small$R|1+\tilde\gamma_2\gamma_2|^2\tilde x_1+D|1+\tilde\gamma_1\gamma_2|^2\tilde x_2
                      -RD|\tilde\gamma_1-\tilde\gamma_2|^2 x_2$}}
                     {\displaystyle{|1+R\gamma_2\tilde\gamma_2+D\tilde\gamma_1\gamma_2|^2}}
\nonumber\,,
\end{eqnarray}
where we omitted the superscripts. Inserting these
expressions into Eq. (\ref{GreenK}), and performing the analogous
calculation for $\hG^K_-$, we obtain explicit expressions for the
phase-sensitive thermal conductance, defined by
$I_{\epsilon}=-\kappa(\phi,T)\;\delta T$, in the limit $\delta
T\rightarrow 0$. The general result can also be used to calculate the
heat current beyond the linear response limit. A more detailed derivation and discussion
will be presented elsewhere \cite{zha03b}.

In the clean limit, $\varepsilon^R\rightarrow\epsilon+i0^+$, the thermal conductance is expressed in
terms of the transmission of bulk excitations of energy $\epsilon\ge\Delta$ and group velocity,
$v_g(\epsilon)=v_f\sqrt{\epsilon^2-\Delta^2}/\epsilon$ through the junction,
\begin{equation}
\kappa(\phi,T)=\Area\int_{\Delta}^{\infty}d\epsilon\,{\cal N}(\epsilon)\left[\epsilon v_g(\epsilon)\right]
\D(\epsilon,\phi)\left(\pder{f}{T}\right)\,,
\label{eq:conductance}
\end{equation}
where $f(\epsilon)=(e^{\epsilon/T}+1)^{-1}$ is the Fermi function, $\cN(\epsilon)=N_f\epsilon/\sqrt{\epsilon^2-\Delta^2}$ is the bulk
density of states (DOS), and $\D(\epsilon,\phi)=\D_{\text{ee}}(\epsilon,\phi)+\D_{\text{eh}}(\epsilon,\phi)$
is the transmission coefficient for the heat current, which is the sum of
\begin{equation}
\D_{ee}(\epsilon,\phi)=D\frac{(\epsilon^2-\Delta^2)(\epsilon^2-\Delta^2\cos^2\frac{\phi}{2})}
                              {\left[\epsilon^2-\Delta^2(1-D\sin^2\frac{\phi}{2})\right]^2}
\,,
\label{eq:D_ee}
\end{equation}
the transmission coefficient for electron-like (hole-like) quasiparticles remaining electron-like (hole-like), and
\begin{equation}
\D_{eh}(\epsilon,\phi)=D(1-D)\frac{(\epsilon^2-\Delta^2)\,\Delta^2\sin^2\frac{\phi}{2}}
                              {\left[\epsilon^2-\Delta^2(1-D\sin^2\frac{\phi}{2})\right]^2}
\,,
\label{eq:D_eh}
\end{equation}
the transmission coefficient for electron-like (hole-like) quasiparticles with branch conversion to hole-like
(electron-like) quasiparticles. Here we neglect the angular dependence of the barrier transmission and reflection
probabilities \footnote{More realistic models for the barrier
transmission which take into account the reduction in the transmission probability for trajectories away from
normal incidence do not lead to significant changes in the results for constant $D$.}.

The heat current density carried by bulk quasiparticles of energy $\epsilon$ in the superconducting leads
reduces to the normal-state current density, $\cN(\epsilon)[\epsilon\,v_g(\epsilon)]=N_f\epsilon\,v_f$;
the increase in bulk DOS is compensated by the reduction in the group velocity.
For $\phi=0$, $\D(\epsilon,\phi)$, reduces to the barrier transmission probability for normal-state quasiparticles,
$\D(\epsilon,0)=D$. Thus, for $\phi=0$ the thermal conductance of the ScS contact is simply reduced by the opening of
the gap in the quasiparticle spectrum.  However, for $\phi\ne 0$ the transmission coefficient, $\D(\epsilon,\phi)$, includes
the modification of the local DOS near the contact by the formation of an Andreev bound state (ABS) with energy below the gap
edge, as well as the particle-hole coherence amplitudes which alter direct ($ee$) transmission and generate branch conversion ($eh$) scattering.

\begin{figure}
\includegraphics[height=2.4in]{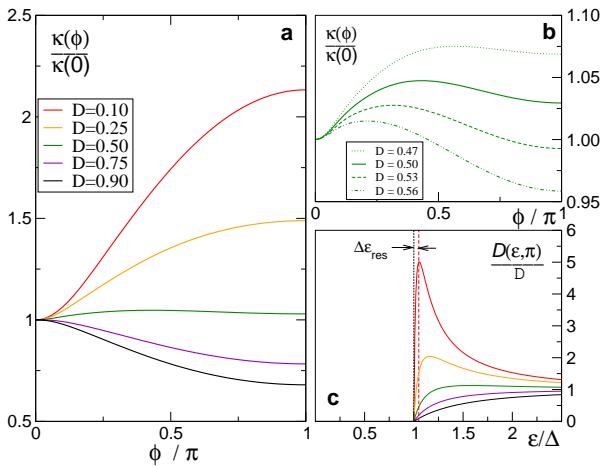}
\caption{a) The thermal current as a function of $\phi$ and barrier transparency, $D$.
            The thermal conductance is normalized for each $D$ by its value at $\phi=0$.
         b) Non-monotonic oscillations of $\kappa(\phi)$ for $D\simeq 0.5$.
         c) Normalized transmission coefficient at $\phi=\pi$ showing the resonant transmission
            for $D\ll 1$, and the suppressed transmission for $D\rightarrow 1$. The resonance peak
        is at $\epsilon_{\text{res}}\simeq\Delta(1+D/2)$.}
\label{fig:phase-modulation}
\end{figure}

The relative importance of the direct ($ee$) and branch-conversion ($eh$) processes to the phase dependence of
the thermal conductance depends on the barrier transparency $D$. For both processes the ABS plays a central role
in controlling the phase modulation of the conductance. The bound state leads to a reduction in the local DOS near
the contact and to a corresponding suppression of the transmission coefficient for excitations with
$\epsilon=\Delta$. For moderate to high transmission barriers ($D\gtrsim 0.5$) this leads to suppression of the
thermal conductance for $\phi\approx\pi$. For low transmission barriers ($D\ll 1$) multiple Andreev reflection leads
to a shallow bound state just below the continuum edge at $\epsilon_{\text{b}}=\Delta\sqrt{1-D\sin^2(\frac{\phi}{2})}$.
The spectral weight of the ABS is derived from the continuum states near $\epsilon=\Delta$, which suppresses the divergence
at $\epsilon=\Delta$ at the cost of a large, but finite, resonant enhancement in the transmission of quasiparticles at
energies, $\epsilon\approx\Delta(1+\tinyonehalf D\sin^2(\frac{\phi}{2}))$, {\sl above} the gap (Fig. \ref{fig:phase-modulation}c).
The resonance generates a strong enhancement of the thermal conductance as the phase is tuned to $\phi=\pi$. These features, as
well as the evolution of the phase-modulation of the conductance with barrier transmission, are shown in Fig. \ref{fig:phase-modulation},
where we plot the normalized conductance, $\kappa(\phi)/\kappa(\phi=0)$ for $0 < D \le 1$ and $T=0.5\Delta$ ($T=0.72\,T_c$).
For intermediate values of the barrier transparency, $D\simeq 0.5$, the phase dependence of the conductance is a
non-monotonic function of $\phi$ for $0\le\phi\le\pi$ (Fig. \ref{fig:phase-modulation}b), although the amplitude of these
oscillations is very small. For $D=1$, transmission with branch conversion drops out ($\D_{\text{eh}}=0$) and the transmission
coefficient for the quasiparticle heat current reduces to $\D(\epsilon,\phi)=(\epsilon^2-\Delta^2)/(\epsilon^2-\Delta^2\cos^2(\frac{\phi}{2}))$,
with a resulting thermal conductance in agreement with Ref. \cite{kul92} for a pinhole.

We compare Eq. \ref{eq:conductance} for $D\ll 1$ with the heat current obtained by perturbation theory
from a tunnelling Hamiltonian (tH) description of SIS tunnel junctions. Based on the tH method
Guttman et al. \cite{gut97} obtained a heat current of the form: $I_{0}+I_{1}\cos\phi$.
If we expand $\D$ to leading order in the barrier transmission probability, $D$, we obtain the tH result
for the conductance from the linear response limit of Eq. (2) of Ref. \cite{gut97}b,
\begin{equation}
\hspace*{-2mm}\kappa^{\text{tH}}=\Area N_f v_f\,D\,
                          \int_{\Delta}^{\infty}d\epsilon\,\epsilon
                          \frac{\epsilon^2-\Delta^2\cos\phi}
                               {\epsilon^2-\Delta^2}
                          \left(\pder{f}{T}\right)
                          \,.
\label{eq:tunnel-current}
\end{equation}
This result has an unphysical divergence due to the singularity at $\epsilon=\Delta$. In the tH method the
divergence is regulated by an {\sl ad hoc} procedure. Guttman et al. \cite{gut97} required a finite
temperature difference and $\Delta(T_1)\ne\Delta(T_2)$ for the gaps of the two superconductors. However, Eqs.
\ref{eq:conductance}-\ref{eq:D_eh} for arbitrary transparency show that the unphysical divergence is not a
singularity of the limit $\delta T\rightarrow 0$, but a failure of perturbation theory in the tunnel
Hamiltonian; which does not include the change in the spectrum near the contact. The bound state and the
resonance regulate the singularity obtained in perturbation theory and lead to a thermal conductance that is
nonanalytic, $\sim D\ln D$, but vanishes for $D\rightarrow 0$. The result for $\kappa(\phi)$ to order $D$ also
has non-perturbative corrections to the phase modulation of the conductance, which includes terms
$\propto\cos\phi$ as well as $\sin^2{\phi\over 2}\ln(\sin^2{\phi\over 2})$,
\begin{equation}
\kappa(\phi)=\kappa_0-\kappa_1\sin^2{\phi\over 2}\ln(\sin^2{\phi\over 2})+\kappa_2 \sin^2{\phi\over 2}
\,,
\label{eq:conductance_small_D}
\end{equation}
where $\kappa_1=k\,\sech^2(\Delta/2T)$,
$k=\Area\,N_f v_f D(\Delta^3/4 T^2)$,
$\kappa_0=k\int_{1}^{\infty}dx\,x^2\,\sech^2(x\Delta/2T)$,
$\kappa_2=-(1+\ln D)\,\kappa_1$
$+k[(4T/\Delta)(1-\tanh(\Delta/2T))+c]$, and
$c=2\int_0^{\infty}dx\,x\ln{x}$
$[1+(\Delta/T)\sqrt{x^2+1}\tanh(\Delta\sqrt{x^2+1}/2T)]$
$(x^2+1)^{-3/2}$ $\sech^2(\Delta\sqrt{x^2+1}/2T)$.
The relative magnitude of the non-perturbative correction, $\kappa_1/\kappa_2$, is approximately $25\%$
for $D=0.01$ at $T/T_c=0.8$ and increases with decreasing temperature.

\begin{figure}
\includegraphics[height=2.4in]{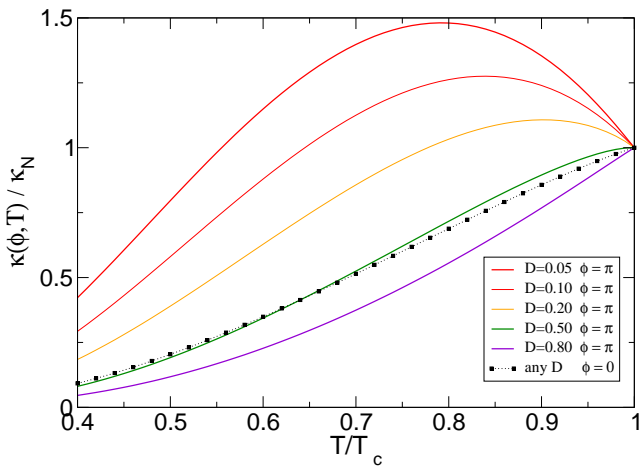}
\caption{The temperature dependence of phase-modulation of the thermal conductance for $\phi=\pi$,
         normalized by the conductance at $T_c$, $\kappa_N=\frac{\pi^2}{6}\Area N_f v_f T_c D$. Shown for
         comparison is the conductance for $\phi=0$.}
\label{fig:conductance-T-dependence}
\end{figure}

For general transparency the magnitude of the phase modulation of the thermal conductance is a maximum for
$\phi=\pi$, except for a small range of barriers with $D\approx 1/2$. The temperature dependence of the
conductance for $\phi=\pi$ is plotted in Fig. \ref{fig:conductance-T-dependence}. The thermal conductance for
$\phi=0$, is also shown for comparison. For moderate to high-transmission junctions ($D \gtrsim 1/2$) the
thermal conductance for $\phi\ne 0$ is suppressed relative to the conductance at $\phi=0$ at all temperatures.
However, for low-transparency junctions ($D\ll 1$) resonant transmission of quasiparticles just above the gap
edge leads to \emph{increase} in the conductance when $T$ drops below $T_c$. This effect is pronounced for
junctions with $D\lesssim 0.2$; its observation would provide a test of this theory of phase-induced resonant
transmission of quasiparticles.

The results above were obtained in the clean limit for the superconducting leads. In
the diffusive limit, $k_f^{-1}\ll\ell\ll\xi_0$, where $\ell$ is the elastic mean-free path,
the excitations and pairing correlations
are governed by Usadel's diffusion equations \cite{usa70}
for the Fermi-surface averaged propagators,
$\hat{\g}_j^{\text{\tiny A,R,K}}=\int d^2\vp_f\hG_j^{\text{\tiny A,R,K}}(\epsilon,\vp_f;\vR)$.
For an ScS contact Nazarov derived a boundary condition for the propagator
in diffusive conductors \cite{naz99},
\begin{equation}
\sigma_2\check{\g}\partial_z\check{\g}\vert_2=\frac{1}{\Area R_{\text{b}}}
\left\langle\frac{2D\left[\check{\g}_2\,,\,\check{\g}_1\right]}
            {4+D\left(\left\{\check{\g}_2\,,\,\check{\g}_1\right\}-2\right)}\right\rangle_D
\,,
\label{eq:dirty_BC}
\end{equation}
where $\check{\g}_j$ is the Keldysh matrix representation for the $\hat{\g}_j^{\text{\tiny A,R,K}}$,
$R_{\text{b}}$ is the barrier resistance for normal leads and $\langle\ldots\rangle_D=\int
dD\rho(D)\ldots/\int dD\,\rho(D)\,D$ is an average over a distribution of channels with transmission
coefficient $D$ characterizing the interface. For a single channel contact with transmission coefficient $D$
application of Eq. \ref{eq:dirty_BC} yields the result from Eqs. \ref{eq:conductance}-\ref{eq:D_eh} obtained
in the ballistic limit with the replacement: $N_f v_f\Area D/4\rightarrow 1/2e^2R_{\text{b}}$. Thus, the phase
modulation of the thermal conductance of small Josephson weak links are the same in the clean and diffusive
limits. This result is due to the cancellation of impurity renormalization of the diagonal and off-diagonal
self-energies in the propagators for s-wave superconductors up to order $a/\xi_{\mbox{\tiny$\Delta$}}$, and
that Nazarov's boundary condition is based on a junction model with a central layer and interface described
by Zaitsev's boundary condition.

In summary, we have presented a theory for  heat transport through Josephson weak links.
For high transmission junctions the reduction in states with $\epsilon \ge \Delta$,
resulting from the formation of an ABS near the Fermi level, leads to a suppression
of the conductance near $\phi=\pi$. For small
transparency, the presence of a shallow bound state produces a
resonance in the continuum just above the gap edge.
This leads to an increase in conductance as the temperature drops below
$T_c$ for junctions with $\phi\approx\pi$. For a single
channel contact, these results are insensitive to impurity scattering and
hold in the clean and dirty limits.

This work was supported by the NSF grant DMR 9972087, and STINT, the Swedish Foundation for International
Cooperation in Research and Higher Education.


\end{document}